\documentclass[11pt,letterpaper]{article}
\usepackage[letterpaper,margin=1in]{geometry}
\usepackage[all,pdf]{xy}
\usepackage{amsfonts}
\usepackage{lscape}
\usepackage{amssymb}
\usepackage{bbm}
\usepackage{color}
\usepackage{amsfonts,amssymb,mathrsfs,amsmath,amssymb,theorem,float}
\usepackage{graphicx}  %% note spelling
\usepackage{multirow}

\newtheorem{theorem}{Theorem}[section]

\newtheorem{cor}[theorem]{Corollary}
\newtheorem{lemma}[theorem]{Lemma}
\newtheorem{alg}[theorem]{Algorithm}
\newtheorem{remark}[theorem]{Remark}
\newtheorem{defi}[theorem]{Definition}

\newtheorem{example}[theorem]{Example}

\newtheorem{assumption}{Assumption}

\def\qed{\hfil {\vrule height5pt width2pt depth2pt}}

\def\qed{\hfil {\vrule height5pt width2pt depth2pt}}
\def\bref#1{(\ref{#1})}

\def\qed{\hfil {\vrule height5pt width2pt depth2pt}}

\def\proof{{\noindent\em Proof.\,\,}}

\def\bref#1{(\ref{#1})}

\def\N{{\mathbb N}}
\def\Z{{\mathbb Z}}
\def\Q{{\mathbb Q}}

\def\F{{\mathbb {F}}}

\def\bref#1{(\ref{#1})}

\def\+{ \oplus}
\def\-{\ominus}
\def\*{\otimes}

\def\deg{\hbox{\rm{deg}}}

\begin{document}
% \clubpenalty=10000
% \widowpenalty = 10000

%\thispagesyle{empty}

%\thanks{\quad Partially
%       supported by a National Key Basic Research Project of China %(2011CB302400) and  by a grant from NSFC (60821002).}}

\title{Sparse Polynomial Interpolation Based on Diversification}
%Faster Deterministic Sparse Interpolation Algorithms for Straight-Line Program Multivariate
\author{Qiao-Long Huang \\
Research center for mathematics and interdisciplinary sciences\\Shandong University, China\\Email: huangqiaolong@sdu.edu.cn}

\maketitle

\abstract
We consider the problem of interpolating a sparse multivariate polynomial over a finite field, represented with a black box. Building on the algorithm of Ben-Or and Tiwari for interpolating polynomials over rings with characteristic zero, we develop a new Monte Carlo  algorithm over the finite field  by doing additional probes.

To interpolate a polynomial $f\in \F_q[x_1,\dots,x_n]$  with a partial degree bound $D$ and a term bound $T$, our new algorithm costs $O^\thicksim(nT\log ^2q+nT\sqrt{D}\log q)$ bit operations and uses $2(n+1)T$ probes to the black box. If $q\geq O(nT^2D)$, it has constant success rate to return the correct polynomial. Compared with previous algorithms over general finite field, our algorithm has better complexity in the parameters $n,T,D$ and is the first one to achieve the complexity of fractional power about $D$, while keeping linear in $n,T$.

A key technique is a randomization which makes all coefficients of the unknown polynomial distinguishable, producing a diverse polynomial. This approach, called diversification, was proposed by Giesbrecht and Roche in 2011.
Our algorithm interpolates each variable independently using $O(T)$ probes, and then uses the diversification to correlate terms in different images. At last, we get the exponents by solving the discrete logarithms and obtain coefficients by solving a linear system.

We have implemented our algorithm in Maple. Experimental results shows that our algorithm can applied to sparse polynomials with large degree. We also analyze the success rate of the algorithm.

\section{Introduction}
Let $f$ be a sparse multivariate polynomial over $\F_q$, the finite filed of size $q$,
\begin{equation}\label{eq-1}
f=\sum_{i=1}^tc_ix_1^{e_{i,1}}\cdots x_n^{e_{i,n}}
\end{equation}
Suppose $f$ is given by a black box $\mathbf{B}_f$, $$\mathbf{B}_f:\F_p^n\rightarrow\F_p$$
On input $(\alpha_1,\dots,\alpha_n)\in \F_p^n$, the black box evaluates and outputs $f(x_1=\alpha_1,\dots,x_n=\alpha_n)$. Given also bounds $D\geq \max_{i=1}^n\deg_{x_i}(f)$ on the partial degree of $f$ and terms bound $T\geq t$. Our goal is to interpolate the polynomial $f$ to obtain the nonzero coefficients $c_i$ and the corresponding exponents $(e_{i,1},\dots,e_{i,n})$ by probing to the black box. Our contribution is as follows.

\begin{theorem}
 Let $f\in\F_q[x_1,\dots,x_n]$, and suppose we are given a black box $\mathbf{B}_f$  which evaluates $f$, an upper bound $D \geq \max_{i=1}^n \deg_{x_j} f$, and an upper bound $T$ on the number of nonzero terms  of $f$.  If $q\geq 2(n+2)T^2D+1$ and a primitive root of $\F_q$ is known, then there exists a probabilistic algorithm which interpolates $f$ with probability at least $3/4$. The algorithm requires $2(n+1)T$ probes and costs $O^\thicksim(nT\log^2 q +nT\sqrt{D}\log q)$ bit operations.
\end{theorem}

\subsection{Background}
Sparse interpolation is an important problem  in computer algebra and symbolic computation.
For example, sparse interpolation is a key part of GCD computation \cite{JM07,MKW05} and also can be applied to the manipulation and factorization of multivariate polynomial and system solving \cite{CKY89,KT90,DK95,DK98}.

The algebraic complexity usually does not coincide with the bit complexity, which also takes into account the potential growth of the actual coefficients in the field $\F$. Nevertheless, in the special, very important, case when the field $\F$ is finite, both complexities coincide up to a constant factor as all the date have the controllable size.
So complexity of interpolation problem over finite field is also very important in the complexity of interpolation.

We now attempt to summarize previously known interpolation algorithms over finite fields.
For the dense representation, one can use the classical method of
Newton/Waring/Lagrange to interpolate in $O^\thicksim(D^n)$ time.

In 1979, Zippel\cite{Z79,Z90} presented the first sparse algorithm in a modern setting. Zippel's algorithm interpolates $f$ one variable at a time, sequentially. This algorithm is probabilistic and its correctness relies heavily on the  Schwartz-Zippel Lemma that if a random evaluation point is chosen from a large enough set, a nonzero polynomial is not zero at this point with high probability. Zippel's algorithm can be used over finite fields $\F_q$. If $q\geq O(nT^2D^2)$, then it will obtain a correct polynomial with a constant success rate. It uses $O(nTD)$ probes to the black box and has cost $O^\thicksim(nTD)$ operation in $\F_q$.

In 1988, Ben-Or and Tiwari \cite{BT88} presented a deterministic algorithm for interpolating a multivariate polynomial over the field with characteristic $0$. The algorithm evaluates the black box at powers of the first $n$ primes; it evaluates at the points $(2^i,3^i,5^i,\dots,p^i_n)$ for $0 \leq i<2T$. Their approach is adapted to finite fields $\F_q$ by using the well-known Kronecker substitution to reduce the multivariate problem to
a univariate one. To interpolate $f(x_1,\dots,x_n)$ with $D>\max_{i=1}^n\deg_{x_i}f$ is the same as to interpolate $\widehat{f}(x,x^D,\dots,x^{D^{n-1}})$. The algorithm requires $q>\deg \widehat{f}=O(D^n)$ and must compute $t$ discrete logarithms to obtain the exponents of $\widehat{f}$. As no polynomial-time algorithm is known for arbitrary finite fields, this will cost $O(tD^{n/2})$ bit operations.

In 1990, Grigoriev et al. \cite{GKS90} proposed a parallel algorithm for sparse polynomial interpolation over finite field $\F_q$. In their algorithm, $q-1$ is the degree bound in each variable. In their algorithm, both probe and bit complexity are polynomial in $q,T$ and $n$.  This provides an efficient solution to the problem when the cardinality of the $\F_q$ is fixed. But, when $\F_q$ grows large, the number of points of evaluation grows with $q$. In our algorithm, the number of evaluation is independent of $q$.

In \cite{HR99}, Huang and Rao developed the first successful adaptation of the Ben-Or/Tiwari approach to finite fields $\F_q$. In their algorithm,  $4t(t-2)\overline{D}^2 + 1$ probes are needed. Here $\overline{D}$ is a bound of the total degree of $f$. As we know $\deg f\leq n\deg_{x_i}f$, $\overline{D}$ is $O(nD)$.
The key novelty of their algorithm is to replace the primes $2,3,5,\dots,p_n$ in Ben-Or/Tiwari by different linear polynomials in $\F_q[y]$. Their algorithm is Las Vegas and has the restriction $q\geq 4t(t-2)\overline{D}^2 + 1$.

In 2010, Javadi and Monagan \cite{JM10} gave a faster way of adaptation of the Ben-Or/Tiwari algorithm to finite field $\F_q$. Their algorithm interpolates each variable independently using $O(t)$ probes so that the main loop of their algorithm can be parallelized. It does $O(nT)$ probes.

Besides the direct methods above, there are reduction-based methods. In 2001, Klivans
 and Spielman gave the first deterministic polynomial time algorithms for sparse interpolation over finite fields with large characteristic. They reduce the multivariate interpolation problem to the univariate interpolation problem. In their Monte Carlo version, it reduces a multivariate into $n$ univariate polynomials with degree $O(nT^2\overline{D})$. In 2014, Arnold and Roche \cite{AR14} present randomized kronecker substitutions. The reduction works similarly to the classical and widely-used Kronecker substitution. Arnold and Roche's algorithm reduces a multivariate into $O(n+\log T)$ univariate polynomials with degree $O(T\overline{D})$.

 As an application, both of the two reductions gave a new algorithm for multivariate interpolation which uses these new techniques along with any existing univariate interpolation algorithm. For finite fields, using the univariate Ben-Or/Tiwari algorithm which requires $2T$ black box probes and $\widetilde{O}(d)$ bit operations, Klivans and Spielman's reduction algorithm gives a multivariate interpolation with $O(nT)$ probes and $O^\thicksim(nT^2\overline{D}^2)$ operation in $\F_q$; Arnold and Roche's reduction algorithm can give a multivariate interpolation with $O^\thicksim(nT)$ probes and $O^\thicksim(nT\overline{D})$ operation in $\F_q$. But in this algorithm, they need to solve a linear system to obtain the exponents which costs $O(n^\omega T)$ bit operations.
In 2019, Huang and Gao \cite{HG19b} revisited  the randomized kronecker substitutions and gave an improvement which deleted the part $O(n^\omega T)$ from the complexity.

\subsection{Supersparse algorithms}
 An information-theoretic lower bound
on the complexity of multivariate sparse interpolation is $\Omega(nT\log D+T\log q)$,
the number of bits used to encode $f$ in \bref{eq-1}. Hence the bit storage size is proportional to the logarithm of the degree $D$. The interpolation algorithm whose complexity is polynomial in $n,T,\log D$, i.e. polynomial in the sparse representation, is called the $supersparse$ algorithm.

So far, there is no supersparse interpolation algorithm for the polynomial given by a black box in general. In our paper, the black box is always limited to the model that can only evaluate in the ground field and outputs the exact values.

For some special finite fields, supersparse algorithm exists. For example, Kaltofen \cite{K10} demonstrates a method for sparse interpolation over $\F_p$ for primes $p$ such that $p-1$ is smooth hence one has a fast discrete logarithm. This is a modification of the Ben-Or/Tiwari algorithm that uses $O(t)$ probes.

If the black box is modified, supersparse algorithm exists. As far as I know, there are currently four such modifications.

The first modification is the $black$-$box$ $numerical\  model$. This model can probe the polynomial on the unit circle and it outputs a value within a given accuracy. In 1995, Mansour \cite{M95} gave a supersparse algorithm for polynomial with integer coefficients. In 2009, Giesbrecht, Labahn and Lee in \cite{GLL09} present a new algorithm for sparse interpolation for polynomials with floating point coefficients which is a numerical adaptation of Ben-Or /Tiwari algorithm. For other work, see \cite{CL08,AM95,GR11}.

The second one is the $modular\ black\ box$. This model is slightly modified from the traditional black box: Given any prime $p$ and any element $\theta$ in $\Z_p$, the modular black box computes the value of the unknown polynomial $f\in\Q[x]$ evaluated at $\theta$ over the field $\Z_p$. In here, both $p$ and $\theta$ are inputs in the evaluation. This is proposed by Giesbrecht and Roche \cite{CL08}. For other work, see \cite{BJ14,KLW90}.

The third one is the $extended\ domain\ black\ box$ which is introduced by D\'{\i}az and Kaltofen \cite{DK98}. This model is capable of evaluating $f(b_1,\dots,b_n)\in \mathbb{E}$ for any $(b_1,\dots,b_n)\in \mathbb{E}^n$ where $\mathbb{E}$ is any extension field of ground field $\mathbb{F}$. That is, we can change every operation in the black box to work over an extension field, usually paying an extra cost per evaluation
proportional to the degree of the extension. Garg and Schost \cite{GS09} first gave a supersparse interpolation algorithm for the straight-line program. Their algorithm computes $f \mod (x^p-1)$ for any $p$. This can be regarded as evaluating $f$ in the extension field which contains $p$-th root and the ground field. More generally, their algorithm works for any rings. Later, a series of later improvement \cite{AGR16,AGR14,AGR13,GR11,HG19} has improved the complexity.

The fourth one is the $derivative\ black\ box$. This model can not only probe $f$, but also can evaluate $\frac{\partial f}{\partial x_k},k=1,\dots,n$ at any given point. It is proposed in \cite{GHS20} and we give this model inspired by \cite{AKP06}. With this model, we give a supersparse algorithm based on Ben-Or and Tiwari algorithm. The key advantage is that in the derivative of $f$, the information of the exponents is contained in the coefficients. For example, if $cm_i=cx_1^{e_1}\cdots x_n^{e_n}$, then the $k$th derivative is $ce_kx_1^{e_1}\cdots x_k^{e_{k-1}}\cdots x_n^{e_n}$.  As in the finite field, the cost of Ben-Or and Tiwari algorithm is dominated by discrete logarithm, this modification replace the solving of discrete logarithm by a division of coefficients which makes the complexity be supersparse.

\subsection{Early termination}
In our algorithm, we assume the terms bound $T$ and degree bound $D$ are inputs.

We notice that bounds for $\overline{D},D$ are sometimes known in advance  and sometimes have to be guessed themselves. The total degree can be guessed efficiently by using sparse interpolation for the univariate polynomial $P(\alpha_1x,\dots,\alpha_nx)$, where $汐=(\alpha_1,\dots,\alpha_n)$ is a random point in $(\F^*_q)^n$. Similarly, the partial degrees $D_k$ are obtained by interpolating $f(\alpha_1,\dots,\alpha_{i-1},x,\alpha_{i+1},\dots,\alpha_n)$.

In \cite{KL03}, Kaltofen and Lee studied black-box sparse interpolation in the case that no bound $T\geq t=\#f $ is given. They developed the technique of early termination, which probabilistically detects when we have enough black-box queries to interpolate $f$. The number of term can be guessed
efficiently by using early termination.

Hao, Kaltofen and Zhi \cite{HKZ16} gave a modified early termination that permits starting the sequence at the point $(1,\dots,1)$, for scalar fields of characteristic $\neq 2$.
\subsection{Related work}
Our algorithm is a modified version of Ben-Or and Tiwari algorithm. We use the same idea to evaluate the polynomial at the powers of some point $(\alpha_1,\dots,\alpha_n)$.

Our algorithm is inspired by Javadi and Monagan \cite{JM10}. In their algorithm, they first evaluation a list of evaluations $f(\alpha^i_1,\dots,\alpha^i_n),i=0,1,\dots,2T-1$. From these values, they obtain the $m_i(\alpha_1,\dots,\alpha_n),i=1,\dots,t$. It is hard to get the exponents of $m_i$ just from $m_i(\alpha_1,\dots,\alpha_n)$.  To solve the problem, they add the number of probes. For example, to compute the exponents of $x_1$, they probe again $f(\alpha'^{i}_1,\dots,\alpha^i_n),i=0,1,\dots,2T-1$ and obtain $m_i(\alpha'_1,\dots,\alpha_n),i=1,\dots,t$. Only the first variable is varied. Let $\Phi(z)=\prod_{i=1}^t(z-m_i(\alpha_1,\dots,\alpha_n))$
and $\Phi'(z)=\prod_{i=1}^t(z-m_i(\alpha'_1,\dots,\alpha_n))$.
As $$\frac{m_i(\alpha_1,\dots,\alpha_n)}{m_i(\alpha'_1,\dots,\alpha_n)}=
(\frac{\alpha_1}{\alpha'_1})^{e_{i,1}}$$
if $r_k=m_i(\alpha_1,\dots,\alpha_n)$ and $\overline{r}_j=m_i(\alpha'_1,\dots,\alpha_n)$, then $r_k=\overline{r}_j(\frac{\alpha_1}{\alpha'_1})^{e_{i,1}}$.

So to determine if $\overline{r}_j(\frac{\alpha_1}{\alpha'_1})^{e_{i,1}}$ is a root of $\Phi(z)$, we may determinate the degree $e_{i,1}$. But to compute all the exponents, the complexity will be $O(nT^2D)$ operation in $\F_q$.

Our algorithm has two main key ingredients different from Javadi and Monagan's. The first one is that we use the information of the coefficient to match the same terms in the different images, which makes the complexity about $T$ from $T^2$ into $T$. To do that we introduce the technique of diversification. We randomly choose $\zeta_1,\dots,\zeta_n$ from $\F^{*n}_q$, with high probability, $f(\zeta_1x_1,\dots,\zeta_nx_n)$ is a polynomial whose different terms have different coefficients. This method is first proposed by Giesbrecht and Roche \cite{GR11}. So our algorithm, instead of returning $m_i(\alpha_1,\dots,\alpha_n),i=1,2,\dots,t$, returns the pairs $(m_i(\alpha_1,\dots,\alpha_n),c_i),i=1,2,\dots,t$. The second ingredient is the evaluation. For the $k$-th variable, we compute $(m_i(\alpha_1,\dots,\alpha_k\omega,\alpha_n),c_i),i=1,2,\dots,t$, where $\omega$ is a primitive of $\F_q$.

As $c_i$'s are different values, we can match the same terms. So

$$\frac{m_i(\alpha_1,\dots,\alpha_k\omega,\dots\alpha_n)}{m_i(\alpha_1,\dots,\alpha_n)}=
\omega^{e_{i,k}}$$

Computing a discrete logarithm, we obtain the $e_{i,k}$. As $e_{i,k}\leq D$, it equals to compute a discrete logarithm within an interval $[0,D]$, which cost $O(\sqrt{D})$ operations in $\F_q$. So our work reduce the complexity about $D$ from $D$ to $\sqrt{D}$.  Compare to Javadi and Monagan's algorithm, we feel that this interpolation algorithim is conceptually much simpler.

\subsection{Compared with previous algorithm}
In this paper, we only consider the problem of interpolation of polynomial given by black box. This is the most general model in the interpolation of polynomial.

In the following algorithms, all of whose computational complexity is polynomial in $T,n,D$ and $\log q$, no supersparse algorithm exists and even no one knows if it exists.

\begin{table}[htp]
\centering
\caption{``Soft-Oh" comparison of interpolation algorithms over finite field $\F_q$ }\label{tab-1}
\scalebox{0.80}[0.80]{%
\begin{tabular}{c|c|c|c|c}
&Probes $(\rho)$ &Bit complexity $(\Theta)$ &Size of $\F_q$&type \\\cline{1-5}
%Dense& $D^n$&deterministic\\
Grigoriev-Karpinski-Singer~\cite{GKS90}&&$n^2T^6\log^2(ntq)+q^{2.5}\log^2q$&&Det\\
Huang-Rao~\cite{HR99}&$T^2\overline{D}$&$(T\overline{D})^8((T\overline{D})^5+\log q)\log^2 q$&$q\geq O(T^2\overline{D}^2)$&LV\\
Javadi and Monagan~\cite{JM10}&$nT$&$T^2(\log q+nD)\log q$&$\phi(q-1)\geq O(nD^2T^2)$&MC\\
Klivans-Spielman~\cite{KS01}&$nT$&$n^2T^2\overline{D}\log q$&$q\geq O(nT^2\overline{D})$&MC\\
Arnold-Roche~\cite{AR14}&$nT$&$nT\overline{D}\log q+n^\omega T$&$q\geq O(T\overline{D})$&MC\\
Huang-Gao~\cite{HG19} &$nT$&$nT\overline{D}\log q$&$q\geq O^\thicksim(T\overline{D})$&MC\\
Zippel~\cite{Z79,Z90} &$nTD$&$nTD\log q$&$q\geq O(nD^2T^2)$&MC\\\cline{1-5}
This paper (The. \ref{the-2}) &$nT$&$nT\sqrt{D}\log q+nT\log^2 q$&$q\geq O(nT^2D)$&MC
\end{tabular}}
\end{table}

``Probes" is the number of evaluations for the polynomials, ``Bit complexity" is the complexity besides the probes, and ``Size of $\F_q$" means that the algorithm can work for
the finite field whose size satisfies this condition, and in the contrary case, the
algorithm need to take values in a proper extension field of $\F_q$. In here, $\overline{D}$ is the total degree bound, so $\overline{D}$ is $O(nD)$.

We can see that our algorithm has better complexity in the parameters $n,T,D$ and our algorithm is the first one to achieve the complexity of fractional power about $D$, while keeping linear in $n,T$. If $q$ is $O^\thicksim((nTD)^{O(1)})$, our algorithm has better complexity than all other methods.
%
%Comparing to Zippel's algorithm, our algorithm has the same bit complexity but needs less evaluations and works for a smaller field.
%
%Actually, our algorithm is the only one which achieves the
%best current bounds in all three parameters in Table \ref{tab-1}.

\subsection{Organization}
Our paper is organized as follows. In Section 2 we present some preliminaries which will be used in our algorithm later.
In Section 3 we present our new algorithm and analyze its complexity and its success rate.
In Section 4 we present an example to show the main row and the key features of our algorithm.
In Section 5 we show the  implementations of our algorithm  on various sets of polynomials.
Finally, a conclusion is given in section 6.

\section{Preliminaries}
\subsection{An order in finite field}\label{sec-2-1}
We will use the coefficients to distinguish the different terms in a polynomial. To speed up the search process, a trivial order in the finite field $\F_q$ is constructed by size comparison of representation elements.

For $q=p^s$, $p$ prime, we suppose that elements of $\F_q$ is represented as $\Z_p[y]/\langle \Phi(y)\rangle$, where $\Phi$ is a irreducible polynomial over $\Z_p$ with degree $s$.  Then we know $\{1,y,\dots,y^{s-1}\}$ is a basis of $\F_q/\Z_p$. For any element $a$ in $\F_q$, there exists unique $(\rho_0,\rho_1,\dots,\rho_{s-1})\in \Z^n_p$ such that $$a=\rho_1+\rho_2y+\cdots+\rho_{s-1}y^{s-1}$$

First we define the order in $\Z_p$. Represent each element of $\F_p$ in the numbers $0,1,\dots,p-1\in \N$. Let $\rho_1,\rho_2\in\F_p$, we say $$\rho_1\preceq \rho_2 \text{ in } \Z_p \text{ iff } \rho_1\leq \rho_2 \text{ in } \N$$
Generalize this order to $(\rho_1,\dots,\rho_n)\in \Z^n_p$ in lexicographic order, then it is an order in $\F_q$.

\subsection{Diversification}

Now we introduce the notion of diversification. If all the coefficients of a polynomial $f$ are all different, then each coefficient can be regarded as the feature of the correlate term. We have the following definition.
\begin{defi}\label{def-1}
If a polynomial $f\in\F_q[x_1,\dots,x_n]$, has all coefficients distinct; that is, $f=\sum_{i=1}^tc_im_i$ and $c_i=c_j\Rightarrow i=j$, then we say $f$ is diverse.
\end{defi}

The diverse polynomials are only special kind in $\F_q[x_1,\dots,x_n]$, which are not the general case in pratice. So we introduce the method of $diversification$ which first introduced by Giesbrecht and Roche \cite{GR11}.

%In our algorithm, the coefficients play an important role in the computing.
%\begin{defi}
%If a polynomial $f\in\F_q[x_1,\dots,x_n]$, has all coefficients distinct; that is, $f=\sum_{i=1}^tc_iM_i$ and $c_i=c_j\Rightarrow i=j$, then we say $f$ is diverse.
%\end{defi}

As we know that if $\zeta_1,\dots,\zeta_n\in\F_q/\{0\}$, then the polynomials
$$f(\zeta_1x_1,\dots,\zeta_nx_n)\leftrightarrow f(x_1,\dots,x_n)$$
 are one-to-one corresponding. And the two polynomials have the same monomials and the only difference between them is the coefficients. So we can interpolate $f(\zeta_1x_1,\dots,\zeta_nx_n)$ instead of $f(x_1,\dots,x_n)$.

If $$f(x_1,\dots,x_n)=\sum_{i=1}^tc_ix^{e_{i,1}}_1\cdots x^{e_{i,n}}_n$$
then $$f(\zeta_1x_1,\dots,\zeta_nx_n)=\sum_{i=1}^tc_i\zeta^{e_{i,1}}_1\cdots \zeta^{e_{i,n}}_nx^{e_{i,1}}_1\cdots x^{e_{i,n}}_n=\sum_{i=1}^t\widetilde{c}_ix^{e_{i,1}}_1\cdots x^{e_{i,n}}_n$$
where $\widetilde{c}_i=c_i\zeta^{e_{i,1}}_1\cdots \zeta^{e_{i,n}}_n$.

Now the coefficients of the new polynomial $f(\zeta_1x_1,\dots,\zeta_nx_n)$ are $\widetilde{c}_i$'s, which depend on the choose of $(\zeta_1,\dots,\zeta_n)$.
Giesbrecht and Roche \cite{GR11} proved that if $(\zeta_1,\dots,\zeta_n)$ are randomly chosen from $\F^{*n}_q$ and $\F_q$ has enough many elements, then $f(\zeta_1x_1,\dots,\zeta_nx_n)$ is diverse with high probability. This is a surprisingly simple but effective trick.

 If the polynomial $f(\zeta_1x_1,\dots,\zeta_nx_n)$ is diverse, then each coefficient of $f(\zeta_1x_1,\dots,\zeta_nx_n)$ is the feature of the correlate term, as it is unique. Once $g=f(\zeta_1x_1,\dots,\zeta_nx_n)$ is known, then $$f=g(\zeta^{-1}_1x_1,\dots,\zeta^{-1}_nx_n)$$

\section{Sparse interpolation over finite fields}
Now we give a sparse interpolation algorithm for black-box multivariate polynomials over finite fields $\F_q$.

Let \begin{equation}\label{eq-2}
f(x_1, \dots, x_n)=c_1m_1+\cdots+c_t m_t\in\F_q[x_1,\dots,x_n]
\end{equation}
be the polynomial to be interpolated,  where $m_i=x_1^{e_{i,
1}}\dots x_n^{e_{i, n}}$ are distinct monomials,  $c_i$ are non-zero
coefficients,  and $t=\#f$ is the number of terms in $f$ and $T\geq t$ is a term bound.

Let $$\beta_i=(\alpha^i_1, \dots, \alpha^i_n)\in \F_q^{*n},\zeta=(\zeta_1, \dots, \zeta_n)\in \F_q^{*n}$$
Denote $$\zeta*\beta_i=(\zeta_1\alpha^i_1,\dots,\zeta_n\alpha^i_n)$$

Now evaluate the value
$$a_i=f(\zeta*\beta_i)=f(\zeta_1\alpha^i_1,\dots,\zeta_n\alpha^i_n),i=0,1,\dots,2T-1$$ and denote $$v_j=m_j(\alpha_1,\dots,\alpha_n),j=1,2,\dots,t$$

This section is organized as follows. In Section \ref{sec-3-1}, we present a diverse version of Ben-Or and Tiwari algorithm. This algorithm will return ${(\widetilde{c}_i,v_i),i=1,\dots,t}$.
In Section \ref{sec-3-2}, we show how to determine the exponents and coefficients by adding probes.
In section \ref{sec-3-3}, we present our new interpolation algorithm and analyse its  time complexity.
In section \ref{sec-3-4}, we analyse the success rate of our new algorithm.

\subsection{The diverse version of Ben-Or and Tiwari algorithm}\label{sec-3-1}

In order to proceed our algorithm, we give the first assumption.
\begin{assumption}
$f$ is a polynomial as in \bref{eq-2}. Let $(\alpha_1,\dots,\alpha_n)\in \F^{*n}_q$ and it satisfies
$$m_i(\alpha_1,\dots,\alpha_n)\neq m_j(\alpha_1,\dots,\alpha_n), \text{ if } i\neq j$$
\end{assumption}

In the following description, we suppose Assumption 1 is correct. First construct an auxiliary polynomial $\Lambda(z)$ to compute the $v_i$. It is constructed as follows.

 \begin{eqnarray}
   \Lambda(z)=\prod_{i=1}^{t} (z-v_i) =z^t+\zeta_{t-1}z^{t-1}+\dots+\zeta_1 z+\zeta_0.
\end{eqnarray}

%Denote $$\widetilde{c}_j=c_jm_j(\zeta_1,\dots,\zeta_n)$$
For any $j\in\N$, we have
\begin{align*}
  a_j=f(\zeta*\beta_j)  &= \sum_{i=1}^tc_im_i(\zeta_1\alpha^j_1,\dots,\zeta_n\alpha^j_n)  \\
     &= \sum_{i=1}^tc_im_i(\zeta_1,\dots,\zeta_n)m_i(\alpha^j_1,\dots,\alpha^j_n)\\
     &= \sum_{i=1}^t\widetilde{c}_im_i(\alpha^j_1,\dots,\alpha^j_n)=\sum_{i=1}^t\widetilde{c}_iv^j_i
\end{align*}

Consider the sum
$$
\sum_{i=1}^{t} \widetilde{c}_i v_i^j \Lambda(v_i)=\sum_{k=0}^{t-1} \zeta_k(\widetilde{c}_1 v_1^{k+j}+\widetilde{c}_2 v_2^{k+j}+\dots+\widetilde{c}_t v_t^{k+j})+(\widetilde{c}_1 v_1^{t+j}+\widetilde{c}_2 v_2^{t+j}+\dots+\widetilde{c}_t v_t^{t+j})
$$
for $j=0, \dots, t-1$. Since $\zeta(v_i)=0$,
$$a_j \zeta_0+a_{j+1} \zeta_1+\dots+a_{j+t-1}\zeta_{t-1}+a_{j+t}=0,0\leq j\leq t-1.$$

We now have the Toeplitz system $A_t\overrightarrow{\lambda}_t=\overrightarrow{b}_t$  where

{\small
\begin{equation}\label{eq-toep}
A_i=\left(\begin{array}{cccc}
a_{t-1}&a_t&\cdots&a_{t-2+i}\\
a_{t-2}&a_{t-1}&\cdots&a_{t-3+i}\\
\vdots&\vdots&\ddots&\vdots\\
a_{t-i}&a_{t-i+1}&\cdots&a_{t-1}\\
\end{array}\right),
\overrightarrow{\lambda}_i=\left(
\begin{array}{c}
\lambda_0 \\
\lambda_1 \\
\vdots \\
\lambda_{i-1}\\
\end{array}
\right),
\overrightarrow{b}_i=-\left(
\begin{array}{c}
a_{t-1+i} \\
a_{t-2+i} \\
\vdots \\
a_t\\
\end{array}
\right)
\end{equation}}

$A_t$ is non-singular as can be seen from the factorization.

%
%This system is non-singular as can be seen from the factorization.
%
{\small \begin{eqnarray}\label{eq-vand2}
A_t&=&\left(\begin{array}{cccc}
v_1^{t-1}&v_2^{t-1}&\cdots&v_t^{t-1}\\
v_1^{t-2}&v_2^{t-2}&\cdots&v_t^{t-2}\\
\vdots&\vdots&\ddots&\vdots\\
1&1&\cdots&1\\
\end{array}\right)
\left(
\begin{array}{cccc}
\widetilde{c}_1&0&\cdots&0 \\
0&\widetilde{c}_2&\cdots&0 \\
\vdots&\vdots&\ddots&\vdots \\
0&0&\cdots&\widetilde{c}_t\\
\end{array}
\right)\left(
\begin{array}{cccc}
1&v_1&\cdots&v_1^{t-1} \\
1&v_2&\cdots&v_2^{t-1}\\
\vdots&\vdots&\ddots&\vdots \\
1&v_t&\cdots&v_t^{t-1}\\
\end{array}
\right)
\end{eqnarray}}

Since the $v_i$ are distinct, the two Vandermonde matrices are nonsingular and as no $\widetilde{c}_i$ is zero, the diagonal matrix is nonsingular, too. If the input value of the upper bound $T$ is greater than $t$, then the coefficients $\widetilde{c}_k$, for $k>t$, can be regarded as zero and the resulting system $A_{T}$ would be singular.

The root of the polynomial $\Lambda(z)$ give the $v_i$ and by choosing the first $t$ evaluations of $f$, we get the following transposed Vandermonde system of equations $V\overrightarrow{\widetilde{c}}=\overrightarrow{a}$ for the coefficients of $f$, where

{\small
\begin{equation}\label{eq-vand}
V=\left(\begin{array}{cccc}
1&1&\cdots&1\\
v_1&v_2&\cdots&v_t\\
\vdots&\vdots&\ddots&\vdots\\
v_1^{t-1}&v_2^{t-1}&\cdots&v_t^{t-1}\\
\end{array}\right),
\overrightarrow{\widetilde{c}}=\left(
\begin{array}{c}
\widetilde{c}_1 \\
\widetilde{c}_2 \\
\vdots \\
\widetilde{c}_t\\
\end{array}
\right),
\overrightarrow{a}=\left(
\begin{array}{c}
a_0 \\
a_1 \\
\vdots \\
a_{t-1}\\
\end{array}
\right)
\end{equation}}

We now state the algorithm for computing the pairs $(\widetilde{c}_i,v_i)$.
The key  ingredients are:
\begin{itemize}
\item Compute $v_1,\dots,v_t$ from the $b_0,b_1,\dots,b_{2T-1}$.
\item Compute the coefficients $\widetilde{c}_1,\dots,\widetilde{c}_t$ from $b_0,b_1,\dots,b_{t-1}$ and $v_1,\dots,v_t$.
\end{itemize}

\begin{alg}[Monomials and coefficients(MC)]\label{alg-1}
\end{alg}

{\noindent\bf Input:}
\begin{itemize}
\item A black box $\mathbf{B}_f:\F_q^n\rightarrow \F_q$  where $f\in \F_q[x_1,\dots,x_n]$ is the target polynomial.
\item A point $(\alpha_1,\dots,\alpha_n)\in \F^{*n}_q$ that satisfies  Assumption 1.
\item A point $(\zeta_1,\dots,\zeta_n)\in \F^{*n}_q$.
\item A terms bound $T\geq \#f$.
\end{itemize}

{\noindent\bf Output:} The pairs $\{(\widetilde{c}_i, v_i)|i=1,2,\dots,t\}$ where $\widetilde{c}_1\preceq\cdots\preceq \widetilde{c}_t$.
\begin{description}
\item[Step 1:] For $i=0,1,\dots,2T-1$, probe $a_i:=f(\zeta*\beta_i)=f(\zeta_1\alpha^i_1,\dots,\zeta_n\alpha^i_n)$.

\item[Step 2:] Find the rank $t$ of the matrix $A_T$.

\item[Step 3:] Solve the Toeplitz system $A_t\overrightarrow{\lambda}_t=\overrightarrow{b}_t$ described in \bref{eq-toep} to recover the auxiliary polynomial $\Lambda(z)$.

\item[Step 4:] Compute the roots of $\Lambda(z)$ and denote them $v_1,\dots,v_t$.

\item[Step 5:] Find the coefficients $\widetilde{c}_i$ by solving the transposed Vandermonde system $V\overrightarrow{\widetilde{c}}=\overrightarrow{a}$ described in \bref{eq-vand}.

\item[Step 6:] Sort $(\widetilde{c}_1,\dots,\widetilde{c}_t)$ into $(\widetilde{c}_{\sigma(1)},\dots,\widetilde{c}_{\sigma(t)})$ such that $\widetilde{c}_{\sigma(1)}\preceq\dots \preceq \widetilde{c}_{\sigma(t)}$, where $\sigma$ is a permutation of $1,2,\dots,t$.

\item[Step 7:] Return $\{(\widetilde{c}_{\sigma(i)},v_{\sigma(i)})|i=1,2,\dots,t\}$.

 \end{description}

\begin{lemma}\label{lm-4}
Algorithm \ref{alg-1} is correct and  it needs $2T$ evaluations of $f$ plus $O^\thicksim(T\log^2 q)$ bit operations.
\end{lemma}
\proof
The correctness comes from the early description. Now we analyse the complexity.
Due to the fast integer and polynomial multiplication algorithms \cite[p.232]{GG99}, one can perform an arithmetic operation in $\F_{q}$ in $O^\thicksim(\log q)$ bit operations.

In Step 1, it needs $2T$ evaluations.

In Step 2 and Step 3, it needs $O(M(t)\log t\log q)$ bit operations \cite{KY88}.

In Step 4, by \cite[Cor.14.16]{GG99}, it needs expected $O^\thicksim(T\log q)$ $\F_q$-operations to compute all $v_1,\dots,v_t$.

In Step 5, it needs $O(M(t)\log t\log q)$ bit operations \cite{KY88}.

In Step 6, as described in Section \ref{sec-2-1}, sorting $(\widetilde{c}_1,\dots,\widetilde{c}_t)$ is the same as sorting $t$ $s$-tuples in lexicographic order. By quick sort, it needs $O^\thicksim(st)$ operation in $\F_p$. So it needs $O^\thicksim(st\log p)=O^\thicksim(t\log q)$ bit operations.

So the total complexity of the algorithm is  $O^\thicksim(T\log^2 q)$  bit operations.\qed

\begin{remark}
Algorithm \ref{alg-1} only returns the $v_i$ and the corresponding coefficients $\widetilde{c}_i$. In the original Ben-Or and Tiwari Algorithm over field of characteristic 0, $v_i$ has the form $v_i=p_1^{e_{i, 1}}\dots p_n^{e_{i, n}}$ where $p_1,\dots,p_n$ are different primes. One can obtain the exponents $e_{i,1},\dots,e_{i,n}$ from the factorization of $v_i$. But Algorithm \ref{alg-1} works over the finite field,
it is difficult to find the exponents from $v_i=\alpha_1^{e_{i, 1}}\dots \alpha_n^{e_{i, n}}$, which is a multi-variate discrete logarithm problem. So Algorithm \ref{alg-1} is only a intermediate algorithm and will be called in the following interpolation algorithm.
\end{remark}

\subsection{Determine the exponents and coefficients}\label{sec-3-2}
In the above sub-section, we only obtain the $v_i(=\alpha^{e_{i,1}}_1\cdots \alpha^{e_{i,n}}_n)$ and $\widetilde{c}_i(=c_i\zeta^{e_{i,1}}_1\cdots \zeta^{e_{i,n}}_n)$. The information we really want is $e_{i,1},\dots,e_{i,n}$ and $c_i$.
We see $$c_i=\frac{\widetilde{c}_i}{\zeta^{e_{i,1}}_1\cdots \zeta^{e_{i,n}}_n}$$
 As $\zeta_1,\dots,\zeta_n$ are known, once we know $e_{i,1},\dots,e_{i,n}$, then $c_i$ can be computed quickly.

So the first thing we have to do is  compute the exponents.
We will do $2nT$ extra probes to do it. For each $x_k$, we use $2T$ more evaluations to determine the exponents in this variable in  the monomials $m_i$.  The reader may find that computing each variable are independent tasks which can therefore be done in parallel

Fix $\omega$ be a generator of $\F^*_q$. Consider the $k$-th variable $x_k$. Let
$$\beta_{i,k}=(\alpha^i_1, \dots,(\alpha_k\omega)^i,\dots, \alpha^i_n)\in \F_q^{*n}$$
Note that we evaluate the $k$-th variable at powers of $\alpha_k\omega$ instead of $\alpha_k$, and the others do not change.

Denote $v_{j,k}=m_j(\alpha_1,\dots,\alpha_k\omega,\dots,\alpha_n)$. Compared with $v_j$, we have

$$\frac{v_{j,k}}{v_j}=\frac{m_j(\alpha_1,\dots,\alpha_k\omega,\dots,\alpha_n)}
{m_j(\alpha_1,\dots,\alpha_n)}=\omega^{e_{j,k}}\in\F_q$$
where $e_{j,k}$ is the $k$-the exponent in the monomial $m_j$.

From  $v_{j,k}$, $v_j$ and $\omega$, we compute the exponent $e_{j,k}\mod (q-1)$ from the discrete logarithm problem.

As $D\geq \max_{i=1}^n\deg_{x_i} f$ is the partial degree bound, $D\geq e_{j,k}$. If $q-1> D$, then the represent element of $e_{j,k}\mod (q-1)$ is $e_{j,k}$ itself.

To compute the $v_{j,k},i=1,2,\dots,t,k=1,2,\dots,n$, we introduce the second assumption.
\begin{assumption}
$f$ is a polynomial as in \bref{eq-2} and $\omega$ is a generator of $\F^*_q$. Let $(\alpha_1,\dots,\alpha_n)\in \F^{*n}_q$ and for any $k=1,2,\dots,n$, it satisfies
 $$m_i(\alpha_1,\dots,\alpha_k\omega,\dots,\alpha_n)\neq m_j(\alpha_1,\dots,\alpha_k\omega,\dots,\alpha_n), \text{if } i\neq j$$
\end{assumption}

We introduce the Assumption 2 because under such assumption,
for any fix $k$, we can compute $v_{1,k},\dots,v_{t,k}$ from the evaluation $$f(\zeta*\beta_{i,k})=f(\zeta_1\alpha^i_1, \dots,\zeta_2(\alpha_k\omega)^i,\dots, \zeta_n\alpha^i_n),i=0,1,\dots,2T-1$$
by Algorithm \ref{alg-1}.

As we can see,  $\widetilde{c}_i$'s are only related the coefficients $c_i$ and the selection of $\zeta_1,\dots,\zeta_n$, so they are independent of $v_j$'s and $v_{j,k}$'s.

To compute $e_{j,k}$, we have to find the $v_{j,k},v_j$ which  corresponds to the same monomial $m_j$.  As both $v_{j,k},v_j$  corresponds to the coefficient $\widetilde{c}_j$ in the output of Algorithm \ref{alg-1}, if all $\widetilde{c}_i$'s are different, we can match them by searching.

So we introduce the third assumption.
\begin{assumption}
$f$ is a polynomial as in \bref{eq-2}. Let $(\zeta_1,\dots,\zeta_n)\in \F^{*n}_q$ and  it satisfies
$$\widetilde{c}_i\neq \widetilde{c}_j, \text{if } i\neq j$$
\end{assumption}

As $$f(\zeta_1x_1,\dots,\zeta_nx_n)=\widetilde{c}_1m_1+\cdots+\widetilde{c}_tm_t$$

The Assumption 3 means that $f(\zeta_1x_1,\dots,\zeta_nx_n)$ is a diverse polynomial according to Definition \ref{def-1}.

\subsection{Algorithm}\label{sec-3-3}

We now state the interpolation. The key  ingredients are:
\begin{itemize}
\item Compute $\{(c_i,v_i),i=1,\dots,t\}$ from the evaluations $f(\zeta_1\alpha^i_n,\dots,\zeta_n\alpha^i_n),i=0,1,\dots,2T-1$ by Algorithm \ref{alg-1}, where $\widetilde{c}_1\preceq\cdots \preceq \widetilde{c}_n$.
\item For each variable $x_k$, compute $\{(c_{i,k},v_{i,k}),i=1,\dots,t\}$ from the evaluations $f(\zeta_1\alpha^i_n,\dots,\zeta_k(\alpha_k\omega)^i,\dots,\zeta_n\alpha^i_n),i=0,1,\dots,2T-1$ by Algorithm \ref{alg-1}, where $\widetilde{c}_{1,k}\preceq\cdots \preceq \widetilde{c}_{n,k}$.
\item For the exponent of $k$-th variable in the monomial $m_i$, compute $e_{i,k}$ from discrete logarithm problem $\frac{v_{k,i}}{v_i}=\omega^{e_{i,k}}$ in $\F_q$.
\item For each coefficient, compute $c_i=\frac{\widetilde{c}_i}{\zeta^{e_{i,1}}_1\cdots \zeta_n^{e_{i,n}}}$.
\end{itemize}

\begin{alg}[Interpolation]\label{alg-2}
\end{alg}

{\noindent\bf Input:}
\begin{itemize}
\item $\mathbf{B}_f:\F_q^n\rightarrow \F_q$  where $f\in \F_q[x_1,\dots,x_n]$  and $q\geq 2(n+2)T^2D+1$.
\item A partial degree bound $D\geq \max_{i=1}^n\deg_{x_i}(f)$.
\item A bound $T\geq t$ on the number of terms in $f$.
\item A generator $\omega$ of $\F^*_q$.
\end{itemize}

{\noindent\bf Output:} The polynomial $f=\sum_{i=1}^tc_im_i$ with probability $\geq \frac34$ or Fail.
\begin{description}
\item[Step 1:] Choose $2n$ nonzero elements $\alpha_1,\dots,\alpha_n,\zeta_1,\dots,\zeta_n$ from $\F^*_q$ at random.
\item[Step 2:] Call Algorithm \ref{alg-1}, let $$\{(\widetilde{c}_i,v_i)|i=1,2,\dots,t\}=\mathbf{MC}(\mathbf{B}_f,(\alpha_1,\dots,\alpha_n),(\zeta_1,\dots,
    \zeta_n),T)$$
$/*$ note that in the output of Algorithm $\mathbf{MC}$, $\widetilde{c}_1\preceq\cdots\preceq \widetilde{c}_n$  $*/$

\item[Step 3:] For $k=1,\dots,n$ do  $/*$determine $\deg_{x_k}(m_i)$ for $1\leq i\leq t$$*/$
 \begin{description}
    \item[a:] Call Algorithm \ref{alg-1}, let
    $$\{(\widetilde{c}_{i,k},v_{i,k})|i=1,2,\dots,t\}=\mathbf{MC}(\mathbf{B}_f,(\alpha_1,\dots,\alpha_k\omega,\dots\alpha_n),(\zeta_1,\dots,
    \zeta_n),T)$$
    \item[b:] For $i=1,2,\dots,t$ do

     If $c_{i,k}\neq c_i$ then return failure. end if.

     Compute $e_{i,k}$ from the discrete logarithm problem $\frac{v_{i,k}}{v_i}=\omega^{e_{i,k}}$ in $\F_q$.

    \end{description}
\item[Step 4:] For $i=1,2,\dots,t$,
compute $c_i=\widetilde{c}_i/\zeta^{e_{i,1}}_1\cdots \zeta_n^{e_{i,n}}$.

\item[Step 5:] Return $\sum_{i=1}^tc_ix_1^{e_{i,1}}\cdots x_n^{e_{i,n}}$.

 \end{description}

We first analyse the complexity of Algorithm \ref{alg-2}. Our algorithm will require randomness. We assume we may obtain a random bit with bit-cost $O(1)$.

\begin{theorem}\label{the-2}
Algorithm \ref{alg-2} needs $2(n+1)T$ evaluations of $f$ plus $O^\thicksim(nT\log^2 q+nT\sqrt{D}\log q)$ bit operations.
\end{theorem}
\proof
In Step 1, randomly choosing $2n$ elements cost $O^\thicksim(n\log q)$ bit operations.

In Step 2 and $\mathbf{a}$ of Step 3, it calls $n+1$ times Algorithm $\mathbf{MC}$, by Lemma \ref{lm-4}, it needs $2(n+1)T$ evaluations and $O^\thicksim(nT\log^2 q)$ bit operations.

In $\mathbf{b}$ of Step 3, it totally needs to solve $nt$ discrete logarithms.  A good
approach over an arbitrary  finite field was given in \cite{P78}, where we can take advantage of the fact that
the order of each discrete $\log$ is bounded by $D\geq \deg_{x_i} f$, with a cost of $O(\sqrt{D})$ operations in $\F_q$ for each discrete logarithm. So the complexity is $O^\thicksim(nt\sqrt{D}\log q)$ bit operations.

In Step 4, to compute $\zeta^{e_{i,1}}_1\cdots \zeta_n^{e_{i,n}}$, it costs $O^\thicksim(n\log D\log q)$ bit operations. So it needs $O^\thicksim(nt\log D\log q)$ bit operations.

%In Step 5, it needs $O(M(t)\log t\log q)$ bit operations.

So the total complexity of the algorithm is  $O^\thicksim(nT\log^2 q+nT\sqrt{D}\log q)$  bit operations.\qed

\begin{remark}
The most expensive step in Algorithm \ref{alg-2} is $\mathbf{b}$ Step $3$. Here, we need to solve $nT$ discrete logarithm. But for some special $q$ where $q-1$ has only small prime factor, the discrete logarithm can be computed quickly with $O(\log^2 D)$ bit operations. Unfortunately, no polynomial-time algorithm is known for arbitrary finite fields.
\end{remark}

\subsection{Analysis of success rate}\label{sec-3-4}

Now we analyse the success rate.
As mentioned before, the correct return of Algorithm \ref{alg-2} relies on the Assumption 1,2,3.
First we prove the following lemma.
\begin{lemma}\label{lm-2}
Assume $m_1$ and $m_2$ are two different monomials with each degree no more $D$. Let $c$ be a nonzero element in $\F_q$. Then there exists no more than
$$D(q-1)^{n-1}$$
$n$-tuples $(\alpha_1,\dots,\alpha_n)\in \F^{*n}_q$ such that $m_1(\alpha_1,\dots,\alpha_n)=c\cdot m_2(\alpha_1,\dots,\alpha_n)$.
\end{lemma}
\proof
Let $g$ be a generator of the multiplicative group $\F^*_q$. Then for each $n$-th tuple $(\alpha_1,\dots,\alpha_n)$ we can assign another $n$-th tuple $\overrightarrow{a}=(a_1,\dots,a_n)\in \N$ such that $\alpha_i=g^{a_i}$. The $a_i$ are elements of $\mathbb{Z}/(q-1)$. Assume $c=g^{b}$. If two terms $m_1(\alpha_1,\dots,\alpha_n)=c\cdot m_2(\alpha_1,\dots,\alpha_n)$
then $$m_1(\alpha_1,\dots,\alpha_n)=\alpha^{\overrightarrow{e}_1}=g^{\overrightarrow{a}\cdot\overrightarrow{e}_1}=
g^bg^{\overrightarrow{a}\cdot\overrightarrow{e}_2}=c\cdot m_2(\alpha_1,\dots,\alpha_n)$$
So
$$\overrightarrow{a}\cdot\overrightarrow{e}_1=b+\overrightarrow{a}\cdot\overrightarrow{e}_2\mod (q-1)$$
which is
$$\overrightarrow{a}\cdot(\overrightarrow{e}_1-\overrightarrow{e}_2)=b\mod (q-1)$$

By the proof of Proposition 8 in \cite{Z90}, there are $k(q-1)^{n-1}$ such tuples such that $\overrightarrow{a}\cdot(\overrightarrow{e}_1-\overrightarrow{e}_2)=0\mod (q-1)$, where $k$ is the GCD of the elements of $\overrightarrow{e}_1-\overrightarrow{e}_2$ and $q-1$. So there are at most $k(q-1)^{n-1}$  tuples such that $\overrightarrow{a}\cdot(\overrightarrow{e}_1-\overrightarrow{e}_2)=b\mod (q-1)$. Since $k\leq D$, there are at most $D(q-1)^{n-1}$ tuples causing the two terms take on the same value.
\qed

We give the following theorem which will be used to analyse an upper bound on the probability that Assumption 1,2,3 are satisfied when we randomly choose $\alpha_1,\dots,\alpha_n,\zeta_1,\dots,\zeta_n$ from $\F^*_q$.

\begin{theorem}\label{the-1}
Let $f=c_1m_1+\cdots+c_tm_t\in \F_q[x_1,\dots,x_n]$ be a $n$-variable polynomial. Let $\omega\in\F^*_q$, $T\geq t$ and $D\geq \max_{i=1}^n\deg_{x_i}f$. If $\alpha_1,\dots,\alpha_n,\zeta_1,\dots,\zeta_n$ be elements from $\F^*_q$ chosen at random.
Then with probability  $$\geq1- \frac{(n+2)T(T-1)D}{2(q-1)}$$  the Assumption 1,2,3 are satisfied.
\end{theorem}
\proof
For Assumption 1, it needs to make sure $$m_i(\alpha_1,\dots,\alpha_n)\neq m_j(\alpha_1,\dots,\alpha_n), \text{if } i\neq j$$ For each pair $(i,j)$ with $i\neq j$, by Lemma \ref{lm-2}, it exists at most $D(q-1)^{2n-1}$ $2n$-tuples $(\alpha_1,\dots,\alpha_n,\zeta_1,\dots,\zeta_n)$ such that $m_i(\alpha_1,\dots,\alpha_n)=m_j(\alpha_1,\dots,\alpha_n)$. Since there are at most $\frac{T(T-1)}{2}$ such pairs,  at most $\frac{T(T-1)D(q-1)^{2n-1}}{2}$ tuples in $\F^{*2n}_q$ do not satisfy Assumption 1.

For Assumption 2, it needs to make sure  $$m_i(\alpha_1,\dots,\alpha_k\omega,\dots,\alpha_n)\neq m_j(\alpha_1,\dots,\alpha_k\omega,\dots,\alpha_n), \text{if } i\neq j$$
for any fixed $k$ in $\{1,2,\dots,n\}$. Since $m_i(\alpha_1,\dots,\alpha_k\omega,\dots,\alpha_n)=\omega^{e_{i,k}}m_i(\alpha_1,\dots,\alpha_n)$
and $m_j(\alpha_1,\dots,\alpha_k\omega,\dots,\alpha_n)=\omega^{e_{j,k}}m_i(\alpha_1,\dots,\alpha_n)$, %
it equals to $$m_i(\alpha_1,\dots,\alpha_n)\neq c_{i,j}m_j(\alpha_1,\dots,\dots,\alpha_n)$$ where $c_{i,j}=\omega^{e_{j,k}-e_{i,k}}$. For each pair $(i,j)$ with $i\neq j$, by Lemma \ref{lm-2}, it exists at most $D(q-1)^{2n-1}$ $2n$-tuples $(\alpha_1,\dots,\alpha_n,\zeta_1,\dots,\zeta_n)$ such that $m_i(\alpha_1,\dots,\alpha_n)=c_{i,j}m_j(\alpha_1,\dots,\alpha_n)$. Since there are at most $\frac{T(T-1)}{2}$ such pairs and $k=1,2,\dots,n$,  at most $\frac{nT(T-1)D(q-1)^{2n-1}}{2}$ tuples in $\F^{*2n}_q$ do not satisfy Assumption 2.

For Assumption 3, it needs to make sure $$c_im_i(\zeta_1,\dots,\zeta_n)\neq c_jm_j(\zeta_1,\dots,\zeta_n), \text{if } i\neq j$$ For each pair $(i,j)$ with $i\neq j$, by Lemma \ref{lm-2}, it exists at most $D(q-1)^{2n-1}$ $2n$-tuples $(\alpha_1,\dots,\alpha_n,\zeta_1,\dots,\zeta_n)$ such that $c_im_i(\zeta_1,\dots,\zeta_n)=c_jm_j(\zeta_1,\dots,\zeta_n)$. Since there are at most $\frac{T(T-1)}{2}$ such pairs, at most $\frac{T(T-1)D(q-1)^{2n-1}}{2}$ tuples in $\F^{*2n}_q$ do not satisfy Assumption 3.

In total,  at most
$$\frac{(n+2)T(T-1)D(q-1)^{2n-1}}{2}$$
tuples $(\alpha_1,\dots,\alpha_n,\zeta_1,\dots,\zeta_n)\in\F^{*2n}_q$ do not satisfy at least one of the Assumption 1,2,3.

As there are $(q-1)^{2n}$ different points in $\F^{*2n}_q$,  with probability  $$\geq\frac{(q-1)^{2n}-\frac{(n+2)T(T-1)D(q-1)^{2n-1}}{2}}{(q-1)^{2n}}=
1-\frac{(n+2)T(T-1)D}{2(q-1)}$$  the Assumption 1,2,3 are all satisfied.
\qed

\begin{cor}
Algorithm \ref{alg-2} is correct.
\end{cor}
\proof
As described earlier, once $(\alpha_1,\dots,\alpha_n,\zeta_1,\dots,\zeta_n)$ satisfies Assumption 1,2,3, Algorithm \ref{alg-2} returns the correct polynomial $f$.
In the input of Algorithm, $q\geq 2(n+2)T^2D+1$, so
$$\frac{(n+2)T(T-1)D}{2(q-1)}\leq \frac{(n+2)T^2D}{2(q-1)}\leq\frac{1}{4}$$
By Theorem \ref{the-1}, algorithm returns the correct polynomial with probability $\geq 1-\frac{1}{4}=\frac{3}{4}$.\qed

\begin{remark}
By the analysis of Theorem \ref{the-1}, Algorithm \ref{alg-1} returns the correct polynomial with probability $\geq 1-\frac{(n+2)T(T-1)D}{2(q-1)}$. So for any $0<\varepsilon<1$, if $q\geq \frac{(n+2)T(T-1)D}{2\varepsilon}+1$, it returns the correct polynomial with probability $\geq 1-\varepsilon$.
\end{remark}

\section{An example}
We demonstrate our algorithm in the following example. For ease of calculation, we let $\F_q=\Z_p$ where $p$ is a prime.
Here we use $x,y$ and $z$ for variables instead of $x_1,x_2$ and $x_3$.
\begin{example}
Let $f=91yz^2+91x^2yz+61x^2y^2z+61z^5+1$ and $p = 101$. Given the number of terms $t=5$, the number of variables $n=3$, a degree bound $D=5$ and the black box that computes $f$, we want to find $f$.
\end{example}

The first step is to randomly pick $2n=6$ elements $\alpha_1=5,\alpha_2=59,\alpha_3=78,\zeta_1=34,\zeta=29,\zeta_3=89$ of $\mathbb{Z}^*_p$.

We first show how Algorithm \ref{alg-1} works.
We evaluate the black box at the points $\zeta*\beta_0,\dots,\zeta*\beta_{2t-1}$ where $\zeta*\beta_i = (\zeta_1\alpha^i_1,\zeta_2\alpha^i_2,\dots,\zeta_n\alpha^i_n)$. Thus we make $2t$ probes to the black box. The reason to use random values from $\mathbb{Z}^*_p$ is that it decreases the probability of two distinct monomials having the same evaluation. Let $a_i$ be the output of the black box on input $\zeta*\beta_i$ and let $A=(a_0,\dots,a_{2t-1})$. In this example we obtain
 $$A=(87, 96, 13, 2, 62, 77, 74, 63, 64, 31)$$

Now we find the rank of the matrix $A_T$ and solve the Toeplitz system $A_t\overrightarrow{\lambda}_t=\overrightarrow{b}_t$ to recover the auxiliary polynomial $\Lambda(z)$.

In our example, $$\Lambda(z)=z^5+61z^4+72z^3+10z^2+35z+23$$
The next step is to find the roots of $\Lambda(z)$. We know that this polynomial is the product of exactly $t=5$ linear factors. The roots are $v_1 = 1,v_2 = 2,v_3 = 11,v_4 = 43$ and $v_5 = 84$. Now we need to compute the coefficients $\widetilde{c}_i$. We do this by
solving the linear system of equations $V\overrightarrow{\widetilde{c}}=\overrightarrow{a}$. After solving we obtain $$\widetilde{c}_1=1,\widetilde{c}_2=54,\widetilde{c}_3=50,\widetilde{c}_4=43,\widetilde{c}_5=33$$

Sort in order of $\widetilde{c}_i$,  the output of Algorithm \ref{alg-1} is
$$\{(\widetilde{c}_i,v_i)|i=1,\dots,5\}=\{(1,1),(33,84),(43,43),(50,11),(54,2)\}$$

Now let's look at the main step, which is to determine the degrees of each monomial in $f$ in each variable. Consider the first variable $x$. Let $\omega$ be a generator of $\Z_p$. In this example we choose $\omega=34$.
We choose the evaluation points $\zeta*\beta_{0,1}, \zeta*\beta_{1,1},\dots,\zeta*\beta_{2t-1,1}$ where $\beta_{i,1}=(\omega^i\alpha^i_1,\alpha^i_2,\dots,\alpha^i_n)$. Note that we  evaluated the first variable at powers of $\omega\alpha_1$ instead of $\alpha_1$.
By Algorithm \ref{alg-1}, it returns
$$\{(c_{i,1},v_{i,1})\}=\{(1, 1), (33, 84), (43, 16), (50, 91), (54, 2)\}$$

Since $$v_{i,1}=m_i(\omega\alpha_1,\alpha_2,\dots,\alpha_n)=\omega^{e_{i1}}m_i(\alpha_1,
\alpha_2,\dots,\alpha_n)=\omega^{e_{i1}}v_i$$

To compute $e_{i1}$, we can solve the problem
$$\omega^{e_{i1}}\equiv \frac{v_{i,1}}{v_i}\mod p$$

We use coefficients to find the relevant terms and solve  five discrete logarithm problems.

After computing, we have $$(\frac{v_{i,1}}{v_i}|i=1,2,\dots,t)=(1,1,45,45,1)$$

Computing the discrete logarithms, we have $$e_{11}=0,e_{21}=0,e_{31}=2,e_{41}=2,e_{51}=0$$

 This gives us the degree of each monomial $m_i$ in the variable $x$.

We proceed to the next variable $y,z$, just like the variable $x$. And we list the results in Table \ref{tab-1}.

\begin{table}[htp]
\centering
\caption{Evaluations and outputs of Algorithm \ref{alg-1}}\label{tab-1}
\scalebox{0.80}[0.80]{%
\begin{tabular}{|c|c|c|c|c|c|}
\hline
Evaluation&\multicolumn{5}{|c|}{Output of Algorithm \ref{alg-1}} \\ \cline{2-6}
$i=0,\dots,9$&$\widetilde{c}_1=1$&$\widetilde{c}_2=33$&$\widetilde{c}_3=43$&$\widetilde{c}_4=50$&$\widetilde{c}_5=54$\\\cline{1-6}
$f(\zeta*\beta_i)$&$v_1=1$&$v_2=84$&$v_3=43$&$v_4=11$&$v_5=2$\\\cline{1-6}
$f(\zeta*\beta_{i,1})$&$v_{1,1}=1$&$v_{2,1}=84$&$v_{3,1}=16$&$v_{4,1}=91$&$v_{5,1}=2$\\ \cline{1-6}
$f(\zeta*\beta_{i,2})$&$v_{1,2}=1$&$v_{2,2}=84$&$v_{3,2}=16$&$v_{4,2}=71$&$v_{5,2}=68$\\ \cline{1-6}
$f(\zeta*\beta_{i,3})$&$v_{1,3}=1$&$v_{2,3}=39$&$v_{3,3}=48$&$v_{4,3}=71$&$v_{5,3}=90$\\ \cline{1-6}
\hline
\end{tabular}}
\end{table}

From the information of Table \ref{tab-1}, we can compute the divisions of $v_{i,k}$ and $v_i$ and we list the results in Table \ref{tab-2}.

\begin{table}[htp]
\centering
\caption{The divisions of $v_{i,k}$ and $v_i$ }\label{tab-2}
\scalebox{0.80}[0.80]{%
\begin{tabular}{|c|c|c|c|c|c|}
\hline
&$i=1$&$i=2$&$i=3$&$i=4$&$i=5$\\\cline{1-6}
$\frac{v_{i,1}}{v_1}$&$1$&$1$&$45$&$45$&$1$\\ \cline{1-6}
$\frac{v_{i,2}}{v_2}$&$1$&$1$&$45$&$34$&$34$\\ \cline{1-6}
$\frac{v_{i,3}}{v_3}$&$1$&$69$&$34$&$34$&$45$\\ \cline{1-6}
\hline
\end{tabular}}
\end{table}

We know $e_{i,k}=\log_{\omega}\frac{v_{i,k}}{v_i}$, according to Table \ref{tab-2}, we have $$m_1=1,m_2=z^5, m_3=x^2y^2z,m_4=x^2yz,m_5= yz^2$$

Now all that's left is the coefficients. As $$c_i=\frac{\widetilde{c}_i}{m_i(\zeta_1,\dots,\zeta_n)}$$

after dividing we obtain $c_1=1,c_2=61,c_3=61,c_4=91,c_5=91$ and hence $f=91yz^2+91x^2yz+61x^2y^2z+61z^5+1$ and we are done.

\section{Experimental results}

In this section, practical performances of the interpolation
algorithm over finite fields will be reported.
We implement Algorithm \ref{alg-1} and Algorithm \ref{alg-2} in Maple. The Maple code can be found in \begin{verbatim}
http://github.com/huangqiaolong/Maple-codes
\end{verbatim}

The codes also include some test examples and their running times. The data are collected on a
desktop with Windows system, 3.19GHz Core $i7$-$8700$ CPU, and 16GB RAM memory.

In Algorithm \ref{alg-1}, we use the Berlekamp-Massey algorithm to solve the Toeplitz systems, use the command $Roots$ in Maple to find the roots, and use the command $mlog$ in Maple to solve the discrete logarithm problem.
%See Remark \ref{rem-dl1} for details.

We randomly construct five polynomials over the finite field $\F_q$ within the given terms bound and partial degree bound, then regard them as black-box polynomials
and reconstruct them with the algorithm. The average times are collected. In our testing, we fix $q=140122640051$ and use the primitive element $\omega=2$.

\begin{figure}[ht]
\begin{minipage}[t]{0.9\linewidth}
\centering
\includegraphics[scale=0.40]{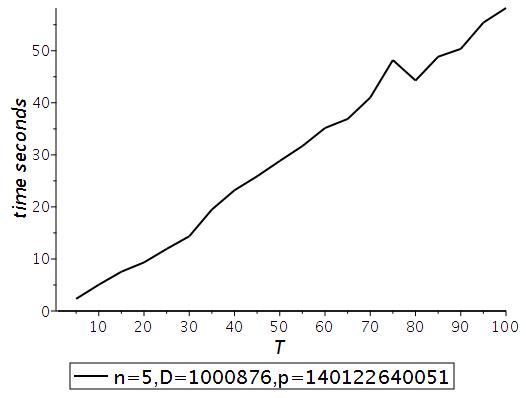}
\caption[KST.jpg]{Average   times with varying $T$} \label{fig1}
\end{minipage}\quad
\end{figure}

\begin{figure}[ht]
\begin{minipage}[t]{0.87\linewidth}
\centering
\includegraphics[scale=0.45]{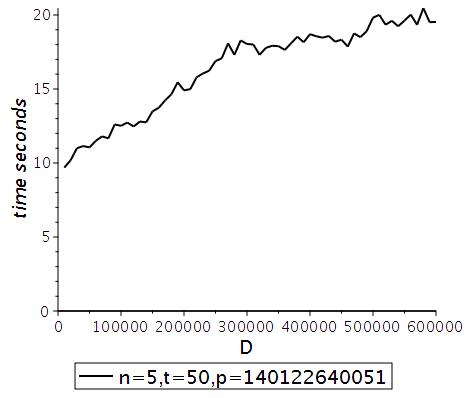}
\caption[KSn.jpg]{Average   times with varying $D$} \label{fig2}
\end{minipage}\quad
\end{figure}

\begin{figure}[ht]
\begin{minipage}[t]{0.95\linewidth}
\centering
\includegraphics[scale=0.40]{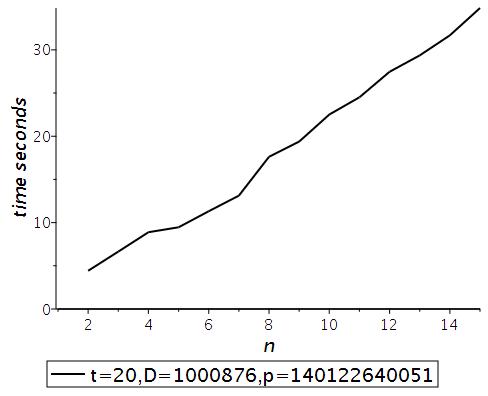}
\caption{Average   times with varying $n$ }\label{fig3}
\end{minipage}
\end{figure}

The results are shown in Figures \ref{fig1}, \ref{fig2}, \ref{fig3}.
In each figure, two of the parameters $n,T,D$ are fixed and one of them is variant.
%
%The relation of the time and the terms is show in Figures \ref{fig1}.
%
%The relation of the time and the number of the variables is show in Figures \ref{fig2}.
%
%The relation of the time and the degree is show in Figures \ref{fig3}.
%
These data are basically in accordance with the complexity
$O^\thicksim(nT\sqrt{D}\log q+nT\log^2 q)$ of the algorithm.

\section{Conclusion}
In this paper, we consider sparse interpolation for a polynomial given by a black box.
The main contribution is a new Monte Carlo algorithm which works over a large finite field .
Our sparse interpolation algorithm is a modification of the Ben-Or/Tiwari algorithm. It does  $2(n+1)T$ probes to the black box. Our algorithm does not interpolate each variable sequentially and thus can more easily be parallelized.  It has lower complexity than any existing algorithms in $n,T,D$. Experimental results show that for sparse polynomials, it does well even if the degree is larger than $1000,000$.

\bibliographystyle{abbrv}
\bibliography{ref}

\begin{thebibliography}{10}

\bibitem{AM95}
N.~Alon and Y.~Mansour.
\newblock $\epsilon$-discrepancy sets and their application for interpolation
  of sparse polynomials.
\newblock {\em Information Processing Letters}, 54(6):337--342, 1995.

\bibitem{AGR13}
A.~Arnold, M.~Giesbrecht, and D.~S. Roche.
\newblock Faster sparse interpolation of straight-line programs.
\newblock In {\em International Workshop on Computer Algebra in Scientific
  Computing}, pages 61--74. Springer, 2013.

\bibitem{AGR14}
A.~Arnold, M.~Giesbrecht, and D.~S. Roche.
\newblock Sparse interpolation over finite fields via low-order roots of unity.
\newblock In {\em Proceedings of the 39th International Symposium on Symbolic
  and Algebraic Computation}, pages 27--34. ACM, 2014.

\bibitem{AGR16}
A.~Arnold, M.~Giesbrecht, and D.~S. Roche.
\newblock Faster sparse multivariate polynomial interpolation of straight-line
  programs.
\newblock {\em Journal of Symbolic Computation}, 75:4--24, 2016.

\bibitem{AR14}
A.~Arnold and D.~S. Roche.
\newblock Multivariate sparse interpolation using randomized kronecker
  substitutions.
\newblock In {\em Proceedings of the 39th International Symposium on Symbolic
  and Algebraic Computation}, pages 35--42. ACM, 2014.

\bibitem{AKP06}
M.~Avenda{\~n}o, T.~Krick, and A.~Pacetti.
\newblock Newton--hensel interpolation lifting.
\newblock {\em Foundations of Computational Mathematics}, 6(1):82--120, 2006.

\bibitem{BT88}
M.~Ben-Or and P.~Tiwari.
\newblock A deterministic algorithm for sparse multivariate polynomial
  interpolation.
\newblock In {\em Proceedings of the twentieth annual ACM symposium on Theory
  of computing}, pages 301--309. ACM, 1988.

\bibitem{BJ14}
M.~Bl{\"a}ser and G.~Jindal.
\newblock A new deterministic algorithm for sparse multivariate polynomial
  interpolation.
\newblock In {\em Proceedings of the 39th International Symposium on Symbolic
  and Algebraic Computation}, pages 51--58. ACM, 2014.

\bibitem{CKY89}
J.~F. Canny, E.~Kaltofen, and L.~Yagati.
\newblock Solving systems of nonlinear polynomial equations faster.
\newblock In {\em Proceedings of the ACM-SIGSAM 1989 international symposium on
  Symbolic and algebraic computation}, pages 121--128. ACM, 1989.

\bibitem{CL08}
A.~Cuyt and W.-s. Lee.
\newblock A new algorithm for sparse interpolation of multivariate polynomials.
\newblock {\em Theoretical Computer Science}, 409(2):180--185, 2008.

\bibitem{MKW05}
J.~de~Kleine, M.~Monagan, and A.~Wittkopf.
\newblock Algorithms for the non-monic case of the sparse modular gcd
  algorithm.
\newblock In {\em Proceedings of the 2005 international symposium on Symbolic
  and algebraic computation}, pages 124--131. ACM, 2005.

\bibitem{DK95}
A.~D{\'\i}az and E.~Kaltofen.
\newblock On computing greatest common divisors with polynomials given by black
  boxes for their evaluations.
\newblock In {\em Proceedings of the 1995 international symposium on Symbolic
  and algebraic computation}, pages 232--239, 1995.

\bibitem{DK98}
A.~D{\i}az and E.~Kaltofen.
\newblock Foxbox a system for manipulating symbolic objects in black box
  representation.
\newblock In {\em Proc. 1998 Internat. Symp. Symbolic Algebraic
  Comput.(ISSAC¡¯98)}, volume~33, pages 30--37, 1998.

\bibitem{GS09}
S.~Garg and {\'E}.~Schost.
\newblock Interpolation of polynomials given by straight-line programs.
\newblock {\em Theoretical Computer Science}, 410(27-29):2659--2662, 2009.

\bibitem{GHS20}
M.~Giesbrecht, Q.-L. Huang, and E.~Schost.
\newblock Sparse multiplication of multivariate linear differential operators.
\newblock {\em Manuscript}.

\bibitem{GLL09}
M.~Giesbrecht, G.~Labahn, and W.-s. Lee.
\newblock Symbolic--numeric sparse interpolation of multivariate polynomials.
\newblock {\em Journal of Symbolic Computation}, 44(8):943--959, 2009.

\bibitem{GR11}
M.~Giesbrecht and D.~S. Roche.
\newblock Diversification improves interpolation.
\newblock In {\em Proceedings of the 36th international symposium on Symbolic
  and algebraic computation}, pages 123--130. ACM, 2011.

\bibitem{GKS90}
D.~Y. Grigoriev, M.~Karpinski, and M.~F. Singer.
\newblock Fast parallel algorithms for sparse multivariate polynomial
  interpolation over finite fields.
\newblock {\em SIAM Journal on Computing}, 19(6):1059--1063, 1990.

\bibitem{HKZ16}
Z.~Hao, E.~L. Kaltofen, and L.~Zhi.
\newblock Numerical sparsity determination and early termination.
\newblock In {\em Proceedings of the ACM on International Symposium on Symbolic
  and Algebraic Computation}, pages 247--254. ACM, 2016.

\bibitem{HR99}
M.-D.~A. Huang and A.~J. Rao.
\newblock Interpolation of sparse multivariate polynomials over large finite
  fields with applications.
\newblock {\em Journal of Algorithms}, 33(2):204--228, 1999.

\bibitem{HG19}
Q.-L. Huang and X.-S. Gao.
\newblock Faster interpolation algorithms for sparse multivariate polynomials
  given by straight-line programs.
\newblock {\em Journal of Symbolic Computation}, 2019.

\bibitem{HG19b}
Q.-L. Huang and X.-S. Gao.
\newblock Revisit sparse polynomial interpolation based on randomized kronecker
  substitution.
\newblock In {\em International Workshop on Computer Algebra in Scientific
  Computing}, pages 215--235. Springer, 2019.

\bibitem{JM07}
S.~M.~M. Javadi and M.~Monagan.
\newblock A sparse modular gcd algorithm for polynomials over algebraic
  function fields.
\newblock In {\em Proceedings of the 2007 international symposium on Symbolic
  and algebraic computation}, pages 187--194. ACM, 2007.

\bibitem{JM10}
S.~M.~M. Javadi and M.~Monagan.
\newblock Parallel sparse polynomial interpolation over finite fields.
\newblock In {\em Proceedings of the 4th International Workshop on Parallel and
  Symbolic Computation}, pages 160--168, 2010.

\bibitem{KLW90}
E.~Kaltofen, Y.~N. Lakshman, and J.-M. Wiley.
\newblock Modular rational sparse multivariate polynomial interpolation.
\newblock In {\em ISSAC}, volume~90, pages 135--139. Citeseer, 1990.

\bibitem{KL03}
E.~Kaltofen and W.-s. Lee.
\newblock Early termination in sparse interpolation algorithms.
\newblock {\em Journal of Symbolic Computation}, 36(3-4):365--400, 2003.

\bibitem{KT90}
E.~Kaltofen and B.~M. Trager.
\newblock Computing with polynomials given byblack boxes for their evaluations:
  Greatest common divisors, factorization, separation of numerators and
  denominators.
\newblock {\em Journal of Symbolic Computation}, 9(3):301--320, 1990.

\bibitem{KY88}
E.~Kaltofen and L.~Yagati.
\newblock Improved sparse multivariate polynomial interpolation algorithms.
\newblock In {\em International Symposium on Symbolic and Algebraic
  Computation}, pages 467--474. Springer, 1988.

\bibitem{K10}
E.~L. Kaltofen.
\newblock Fifteen years after dsc and wlss2 what parallel computations i do
  today: invited lecture at pasco 2010.
\newblock In {\em Proceedings of the 4th International Workshop on Parallel and
  Symbolic Computation}, pages 10--17. ACM, 2010.

\bibitem{KS01}
A.~R. Klivans and D.~Spielman.
\newblock Randomness efficient identity testing of multivariate polynomials.
\newblock In {\em Proceedings of the thirty-third annual ACM symposium on
  Theory of computing}, pages 216--223. ACM, 2001.

\bibitem{M95}
Y.~Mansour.
\newblock Randomized interpolation and approximation of sparse polynomials.
\newblock {\em SIAM Journal on Computing}, 24(2):357--368, 1995.

\bibitem{P78}
J.~M. Pollard.
\newblock Monte carlo methods for index computation (mod $p$).
\newblock {\em Mathematics of computation}, 32(143):918--924, 1978.

\bibitem{GG99}
J.~Von Zur~Gathen and J.~Gerhard.
\newblock {\em Modern computer algebra}.
\newblock Cambridge university press, 1999.

\bibitem{Z79}
R.~Zippel.
\newblock Probabilistic algorithms for sparse polynomials.
\newblock In {\em International Symposium on Symbolic and Algebraic
  Manipulation}, pages 216--226. Springer, 1979.

\bibitem{Z90}
R.~Zippel.
\newblock Interpolating polynomials from their values.
\newblock {\em Journal of Symbolic Computation}, 9(3):375--403, 1990.

\end{thebibliography}

\end{document}